\documentclass[twocolumn]{article}
\usepackage[preprint]{spconfa4}
\usepackage{amsmath,graphicx}
\usepackage{mathtools}
\usepackage{amssymb}
\usepackage{amsfonts}
\usepackage{siunitx}
\usepackage{cite}
\usepackage{url}

\newcommand{\VM}{\boldsymbol}

\newcommand{\MC}{\mathcal}

\newcommand{\NP}{\mathrm{e}}

\newcommand{\T}{\mathsf{T}}
\newcommand{\HT}{\mathsf{H}}

\newcommand{\minimize}{\mathop{\rm minimize}\limits}

\title{Head-Related Transfer Function Interpolation from Spatially Sparse Measurements Using Autoencoder with Source Position Conditioning}
\name{Yuki Ito, Tomohiko Nakamura, Shoichi Koyama, Hiroshi Saruwatari}
\address{Graduate School of Information Science and Technology, The University of Tokyo\\
    7-3-1 Hongo, Bunkyo-ku, Tokyo 113-8656, Japan}
\begin{document}
\ninept
\maketitle
\begin{abstract}

We propose a method of head-related transfer function (HRTF) interpolation from sparsely measured HRTFs using an autoencoder with source position conditioning.
The proposed method is drawn from an analogy between an HRTF interpolation method based on 
regularized linear regression (RLR) and an autoencoder.
Through this analogy, we found the key feature of the RLR-based method that HRTFs are decomposed into source-position-dependent and source-position-independent factors.
On the basis of this finding, we design the encoder and decoder so that their weights and biases are generated from source positions.
Furthermore, we introduce an aggregation module that reduces the dependence of latent variables on source position for obtaining a source-position-independent representation of each subject.
Numerical experiments show that the proposed method can work well for unseen subjects and achieve an interpolation performance with only one-eighth measurements comparable to that of the RLR-based method.
\end{abstract}
        \begin{figure*}[t]
          \centering
          \includegraphics[width=0.9\hsize]{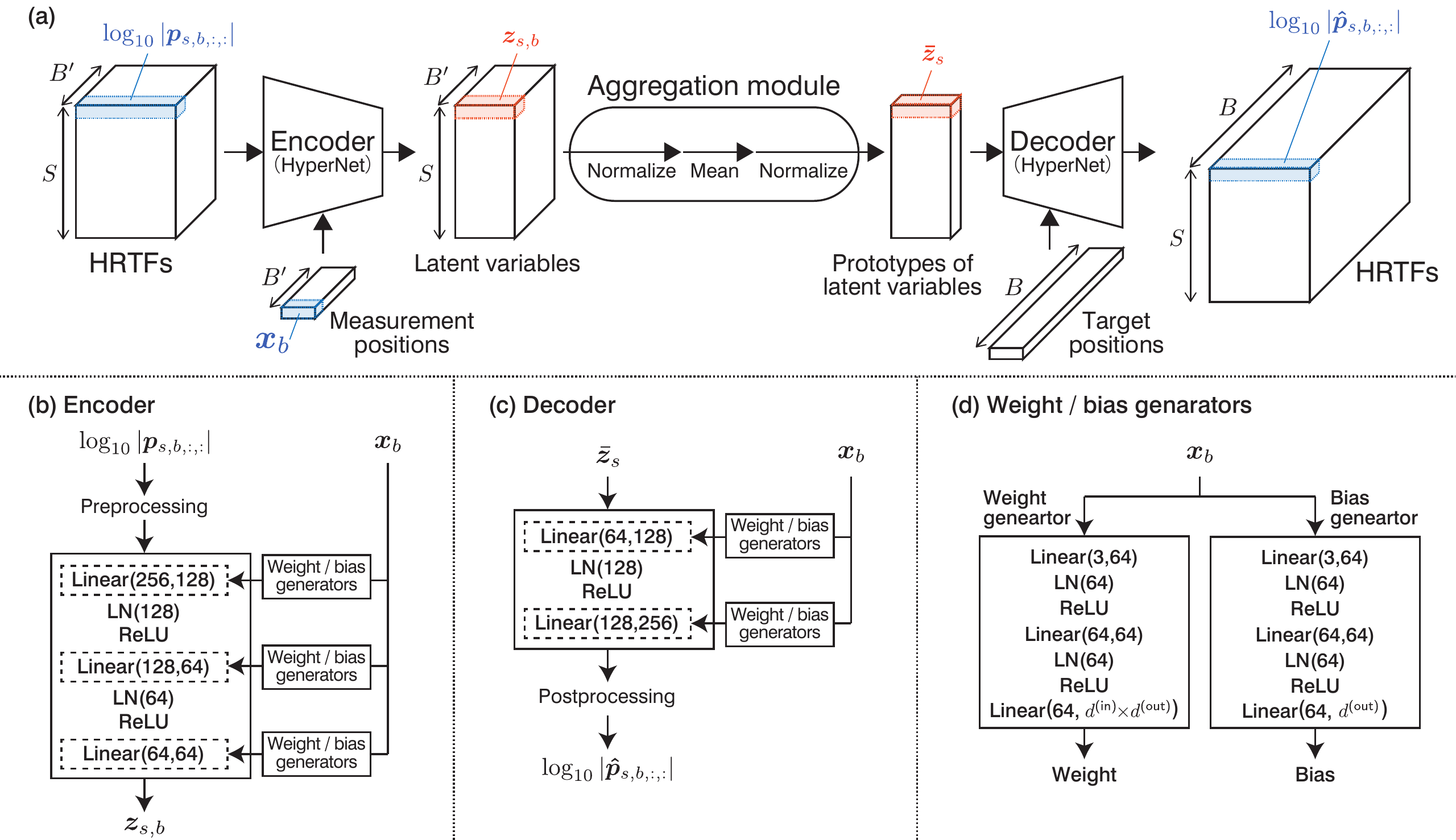} 
          \caption{
          (a) Overview of the proposed model.
          (b) Structure of the encoder.
          The weight/bias generators output the weights and biases for the linear layers enclosed by dashed lines.
          (c) Structure of the decoder. 
          (d) Structure of the weight/bias generators.
          }
          \label{fig:model}
        \end{figure*}
\section{Introduction} \label{sec:intro}
    Recently, demand for high-quality spatial audio technology has increased with the spread of virtual/augmented reality technology. 
    One of the spatial audio methods is the binaural method, which is based on listening with headphones.
    Binaural signals are synthesized by convolving the head-related transfer function (HRTF), which represents the transfer characteristics from the sound source to both ears, into a sound source signal. 
    %
    HRTFs depend on the anatomy of a subject (e.g., pinna shape), which
    vary greatly from subject to subject.
    Hence, using somone else's HRTF leads to poor localization of a sound image~\cite{Wenzel+93}.
    Therefore, the ideal binaural synthesis requires the subject's own HRTFs. 
    However, measuring HRTFs takes a long time because HRTFs are measured by sequentially recording impulse responses from hundreds of source positions in a dense grid on a sphere.
    This measurement typically takes 60 to 90 minutes per subject~\cite{Watanabe+14}.
    Long-time measurements are burdensome for subjects, making it difficult to measure their own HRTFs nonrigorously.
    
    One way to reduce this long measurement time is HRTF interpolation, which refers to spatial upsampling from sparsely measured HRTFs to dense HRTFs.
    %
    The fewer the number of observation positions, i.e., source positions, the shorter the measurement time.
    Typical HRTF interpolation methods use bases in the spatial domain, such as spherical harmonic functions~\cite{Evans+98,Zhang+12,Aussal+13}, spherical wavefunctions~\cite{Duraiswami+04}, and spatial principal components~\cite{Xie12}. 
    %
    As shown in Sect.~\ref{sec:rlr}, these methods amount to solving regularized linear regression (RLR) problems. 
    Thus, when the measured HRTFs are spatially very sparse, we are forced to solve underdetermined problems or 
    to use a small number of bases, 
    which degrades interpolation accuracy, 
    as we will show later in Sect.~\ref{sec:experiment}.
    Therefore, these methods require somewhat dense HRTFs for high-precision interpolation and thus do not contribute to a significant reduction in measurement time.
    
    One approach to overcome this difficulty is a training-based approach.
    The use of deep neural networks (DNNs) has shown promising results for HRTF-related tasks such as real-time binaural synthesis \cite{Bharitkar+18,Tamulionis&Serackis19}, HRTF range extrapolation~\cite{Zhang+19}, and HRTF regression from anthropometric features~\cite{Hu+08,Chun+17,Chen+19,Zhang+20,Xi+21} or ear images~\cite{Lee&Kim18}.
    An HRTF interpolation method using a shallow neural network (NN) has been proposed \cite{Lemaire+05}.
    The network estimates an HRTF magnitude at a target source position from an HRTF magnitude measured at the nearest source position.
    These successes led us to use a DNN.

    
    
    
    We propose a method of interpolating dense HRTF magnitudes from sparse measurements based on an autoencoder conditioned on source positions.
    %
    The proposed method is drawn from our finding that the RLR-based method can be interpreted as an autoencoder, which we will show in Sect.~\ref{sec:proposed}.
    This finding reveals that the key feature of the RLR-based method is to convert measured HRTFs into a subject representation independent of source positions by using source-position-dependent basis expansions.
    Inspired by this feature, we design our network architecture such that only the encoder and decoder, not the latent variables, depend on source positions.
    We also construct a loss function based on cosine distances between the latent variables of each subject at different source positions.
    The proposed model architecture and loss function promote the latent variables to capture the HRTF individuality of each subject.
    

\section{Problem statement} \label{sec:problem}
    As in \cite{Duraiswami+04}, we define HRTF as an acoustic transfer function from the sound source to both ears.
    HRTFs for each ear depend on the physical features of the subject and the position of the sound source.
    %
    Let $p_{s,b,\text{ch},l}\in\mathbb{C}$ denote the HRTF for subject $s\in\{1,\ldots,S\}$, source position $b\in\{1,\ldots,B\}\eqqcolon \MC{B}$, channel $\text{ch}\in\{\text{left},\text{right}\}$, and frequency bin $l\in\{1,\ldots,L\}$, where $S$ is the number of subjects, $B$ is the number of source positions, and $L$ is the number of frequency bins.
    %
    The HRTF interpolation problem of our interest is formulated as
    the problem of estimating spatially dense HRTFs $\{p_{s,b,\text{ch},l}\}_{b\in \MC{B}}$ from spatially sparse HRTFs $\{p_{s,b,\text{ch},l}\}_{b\in \MC{B}^\prime}$, where $\MC{B}^\prime \subset \MC{B}$. Let $B^\prime \coloneqq|\MC{B}^\prime|$.
    To distinguish $b\in\MC{B}^\prime$ from $b\in\MC{B}$, we call the former \textit{measurement position} and the latter \textit{target position}.

    To describe sound source positions, we use three-dimensional Cartesian $\VM{x}_b\coloneqq(x_b,y_b,z_b)$ or spherical coordinates $(r_b,\theta_b,\phi_b)$ with the origin fixed at the center of the subject's head, where $r_b$, $\theta_b$, and $\phi_b$ denote the radius, azimuth angle, and zenith angle, respectively.
    The $+x$ and $+z$ directions of the Cartesian coordinate system are set to the direction of the subject's view and the top of the head, respectively.
    Cartesian coordinates $\VM{x}_b$ and spherical coordinates $(r_b,\theta_b,\phi_b)$ are related by $\VM{x}_b = (r_b\sin\theta_b\cos\phi_b, r_b\sin\theta_b\sin\phi_b,r_b\cos\theta_b)$.

\section{Regularized-linear-regression-based method} \label{sec:rlr}
    We briefly review the HRTF interpolation method based on spherical wavefunction expansions proposed in~\cite{Duraiswami+04}.
    For simplicity, we drop the subscripts $s$ and $\text{ch}$ from $p_{s,b,\text{ch},l}$; thus, we use $p_{b,l}$ in the following.
    On the basis of the reciprocity~\cite{Morse&Ingard86}, an HRTF can be regarded as an acoustic transfer function from a sound source placed at the ear to a microphone placed at an original source position.
    Therefore, the HRTFs can be regarded as the exterior sound field satisfying the homogeneous Helmholtz equation. 
    Such a sound field is approximately represented by the spherical wavefunction expansion as
    \begin{equation}
        p_{b,l} \simeq \sum_{n=0}^{N} \sum_{m=-n}^{n} 
        C_{n,m,l} h_{n}^{(1)}(k_l r_b) Y_{n}^{m}(\theta_b, \phi_b),
        \label{eq:SWFE}
    \end{equation}
    where $h_n^{(1)}(\cdot)$ is 
    the $n$th-order spherical Hankel function of the first kind
    , $Y_{n}^{m}(\cdot)$ is the spherical harmonic function of order $n$ and degree $m$,
    and $(r_b,\theta_b,\phi_b)$ 
    is 
    the original source position 
    in the spherical coordinates~\cite{Williams99}. 
    $k_l=2\pi f_l/v$ is the $l$th wave number, where $f_l$ and $v$ are the $l$th frequency and speed of sound in the air, respectively.
    $\Phi_{n,m,b,l}\coloneqq h_{n}^{(1)}\left(k_l r_b\right) Y_{n}^{m}(\theta_b, \phi_b)$ and $C_{n,m,l}$ are the spherical wavefunction and its expansion coefficient, respectively.
    $N$ is the truncation order and both sides of~\eqref{eq:SWFE} are equal as $N\to\infty$. 
    
    This interpolation method consists of two steps: In the first step, the expansion coefficients $C_{n,m,l}$ are estimated, and in the second step, HRTF $p_{b,l}$ is obtained by substituting the expansion coefficients $C_{n,m,l}$ and 
    the target position
    $(r_b, \theta_b, \phi_b)$ into \eqref{eq:SWFE}.
    The estimation of $C_{n,m,l}$ from HRTFs at 
    $\MC{B}^\prime$
    is as follows.
    We can write \eqref{eq:SWFE} in a matrix--vector product form:
    \begin{equation}
        \VM{p}_{l} \simeq \VM{\Phi}_{l} \VM{c}_{l} \label{eq:SWFE-mat},
    \end{equation}
    where
    $\VM{p}_{l}\in\mathbb{C}^{B^\prime}$ and $\VM{c}_{l}\in \mathbb{C}^{(N+1)^2}$ are vectors consisting of $p_{b,l}$ and $C_{n,m,l}$, respectively, and $\VM{\Phi}_{l}\in \mathbb{C}^{B^\prime\times (N+1)^2}$ is a matrix consisting of $\Phi_{n,m,b,l}$.
    For an overdetermined case,
    i.e., $B^\prime>(N+1)^2$, 
    the estimation problem of $\VM{c}_{l}$ amounts to the following RLR problem:
    \begin{equation}\label{eq:min-prob}
        \minimize_{\VM{c}_{l}\in \mathbb{C}^{(N+1)^2}}\quad\MC{L}_{\text{RLR}} = \left\|\VM{p}_{l} - \VM{\Phi}_{l} \VM{c}_{l} \right\|_2^2 + \lambda
        \left\|\VM{D}^{1/2} \VM{c}_{l}\right\|_2^2,
    \end{equation}
    where $\lambda>0$ is a regularization parameter and $\VM{D}\in\mathbb{R}^{(N+1)^2 \times(N+1)^2}$ is a diagonal matrix whose diagonal components are $1+n(n+1)$. 
    The minimizer of $\MC{L}_{\text{RLR}}$ is given as
    \begin{equation}
            \hat{\VM{c}}_{l} = \left(\VM{\Phi}_{l}^{\HT} \VM{\Phi}_{l} +\lambda \VM{D}\right)^{-1} \VM{\Phi}_{l}^\HT \VM{p}_{l}.
            \label{eq:lininv}
    \end{equation}
    Although only overdetermined cases are discussed in~\cite{Duraiswami+04}, \eqref{eq:lininv} is a reasonable solution for a balanced or an underdetermined case, i.e., $B^\prime\leq(N+1)^2$.

\section{Proposed method} \label{sec:proposed}
    \subsection{Motivation and strategy}
        The RLR-based method described in Sect.~\ref{sec:rlr} works well when a sufficient amount of measured HRTFs of each subject are available.
        However, when measurement positions are spatially sparse, the RLR-based method can be less accurate.
        In fact, as we will show in Sect.~\ref{sec:experiment}, the estimated HRTFs from a small number of measured HRTFs differed from the ground truths in notches and peaks.
        Although decreasing the truncation order $N$ stabilizes the estimation performance, it restricts the expressive power of the spherical wavefunction expansion, which makes it difficult to represent the fine structure of HRTFs.
        
        To overcome this problem, we focus on our finding that the RLR-based method can be reinterpreted from a DNN perspective.
        As described in Sect.~\ref{sec:rlr}, for given $\MC{B}^\prime$ with $\MC{B}$, the RLR-based method consists of two linear transformations given by \eqref{eq:lininv} and \eqref{eq:SWFE-mat}.
        Hence, we can interpret the RLR-based method as a linear autoencoder, where the encoder is a linear layer with weights of $\left(\VM{\Phi}_{l}^{\HT} \VM{\Phi}_{l} +\lambda \VM{D}\right)^{-1}\VM{\Phi}_{l}^\HT$, the decoder is a linear layer with weights of $\VM{\Phi}_{l}$, and the latent variables are the expansion coefficients.
        This analogy reveals the key feature of the RLR-based method.
        That is, it decomposes HRTFs into source-position-dependent and source-position-independent factors, i.e., the spherical wavefunction expansion and expansion coefficients, respectively.
        
        On the basis of this key feature, we propose an HRTF interpolation method using an autoencoder whose encoder and decoder are conditioned on source positions (see Fig.~\ref{fig:model}(a)).
        To obtain a source-position-independent representation, we introduce an aggregation module between the encoder and decoder, which aggregates latent variables of measurement points.
        The proposed autoencoder operates the estimation in the HRTF magnitude domain, similarly to DNN-based methods for HRTF-related tasks~\cite{Chen+19,Xi+21}.
        To obtain the head-related impulse responses (HRIRs) of the estimated HRTF magnitudes, we can use 
        a method of assigning a phase to the HRTF magnitude.
        For example, a method based on minimum phase restoration provides the phase similar to that of the original HRTFs in sound image localization~\cite{Kistler+92}.

    \subsection{Model} \label{sbsec:model}
        \smallskip
        \noindent \textbf{Encoder:}
            The encoder converts the logarithmic magnitudes of measured HRTFs $\left\{\log_{10}\left|\VM{p}_{s,b,:,:}\right|\right\}_{b\in\MC{B}^\prime}$ into the latent variables $\{\VM{z}_{s,b}\}_{b\in\MC{B}^\prime}$, referring to the measurement positions 
            $\MC{B}^{\prime}$
            .
            Here, $|\cdot|$ denotes an element-wise absolute value.
            The measured HRTF is processed independently for each measurement position, which enables us to use the proposed method for various numbers of measurement positions without retraining.
            
            The encoder consists of linear layers conditioned on
            $\MC{B}^{\prime}$
            , layer normalization (LN) layers, and rectified linear unit (ReLU) layers.
            Fig.~\ref{fig:model}(b) shows the encoder architecture, where \textsf{Linear}, \textsf{LN}, and \textsf{ReLU} denote the linear layer, LN layer, and ReLU nonlinearity, respectively.
            The two values in the parenthesis after \textsf{Linear} denote input and output feature sizes, respectively, and the value in the parenthesis after \textsf{LN} denotes a feature size.
            The dimension of latent variables was set to 64.
            For the conditioning on 
            $\MC{B}^{\prime}$
            , we use the hypernetwork idea~\cite{Ha+17}.
            Unlike usual layers, it generates weights and biases of the linear layers from auxiliary information by a DNN, which we call a weight/bias generator.
            Fig.~\ref{fig:model}(d) shows the architecture of the weight/bias generator, where $d^{\textsf{(in)}}$ and $d^{\textsf{(out)}}$ denote the input and output feature sizes of the corresponding linear layer, respectively.
            As an auxiliary information, we can use measurement positions in Cartesian coordinates.
        \smallskip
        \noindent \textbf{Aggregation Module:}
            The aggregation module yields a source-position-independent representation $\bar{\VM{z}}_{s}$, which we call a prototype.
            First, this module normalizes $\VM{z}_{s,b}$ so that its $\ell_2$ norm is one:
            \begin{equation}
                \VM{z}_{s,b}\leftarrow \VM{z}_{s,b}/\|\VM{z}_{s,b}\|_2.
            \end{equation}
            Second, to reduce the dependence of $\VM{z}_{s,b}$ on the source position, the module averages $\VM{z}_{s,b}$ as
            \begin{equation}
                \bar{\VM{z}}_{s} = \frac{1}{B^\prime}\sum_{b\in\MC{B}^{\prime}} \VM{z}_{s,b}.
            \end{equation}
            Note that the averaging operation is inspired by a few-shot learning method presented in \cite{Snell+17}.
            Finally, the prototype is normalized as
            \begin{equation}
                \bar{\VM{z}}_{s} \leftarrow \bar{\VM{z}}_{s}/ \|\bar{\VM{z}}_{s}\|_2.
            \end{equation}
            Since the prototype no longer depends on $b$, we can use it as a representation of a subject.
            
        \smallskip
        \noindent \textbf{Decoder:}
            The decoder converts the prototype into a logarithmic magnitude of the HRTFs $\log_{10}\left|\VM{\hat{p}}_{s,b,:,:}\right|$ at each target position $b\in\MC{B}$.
            Fig.~\ref{fig:model}(c) shows the architecture of the decoder.
            Similarly to the encoder, it consists of two linear layers whose weights and biases are generated from the weight/bias generators, LN layers, and ReLU nonlinearities.
            The inputs of the weight/bias generators are target positions in Cartesian coordinates.
            %
            %
            
    \subsection{Loss function} \label{sbsec:loss}
        To train the proposed model, we use a loss function $\MC{L}$ defined as
        \begin{equation}
            \MC{L} = \text{LSD} + \alpha~ \text{CosDist}\label{eq:loss},
        \end{equation}
        where $\alpha\geq 0$ is a hyperparameter that controls the importance of the second term.
        The first term on the right-hand side of \eqref{eq:loss} is a reconstruction error of HRTFs.
        The second term is expected to make the latent variable $\VM{z}_{s,b}$ less dependent on the 
        source positions
        $\VM{x}_b$ and instead better represent the characteristics of each subject $s$.
        The first term 
        is the log-spectral distortion (LSD) of HRTFs and is defined as
        \begin{equation}
            \text{LSD} \coloneqq \frac{1}{2SB}\sum_{s,b,\text{ch}}
                \sqrt{
                    \frac{1}{L}\sum_{l}\left(
                        20\log_{10}\frac{\left|\hat{p}_{s,b,\text{ch},l}\right|}{\left|p_{s,b,\text{ch},l}\right|}
                    \right)^2
                },\label{eq:LSD}
        \end{equation}
        where $\hat{p}_{s,b,\text{ch},l}$ and $p_{s,b,\text{ch},l}$ denote an estimated and a true element of HRTFs, respectively.
        The second term on the right-hand side of \eqref{eq:loss} 
        is a measure of nonuniformity
        of the latent variables $\VM{z}_{s,b}$ for each subject $s$ and is given as
        \begin{equation}
            \text{CosDist} \coloneqq \sqrt{\frac{1}{SB^{\prime}}
            \sum_{s,b} \left(1-
            \frac{\VM{z}_{s,b}^{\T}\bar{\VM{z}}_{s}}{
            \|\VM{z}_{s,b}\|_2 \| \bar{\VM{z}}_{s}\|_2}
            \right)^2}.
        \end{equation}

\section{Experimental evaluation} \label{sec:experiment}
    
    \subsection{Experiment condition} \label{sbsec:condition}
    \subsubsection{Preparation of training, validation, and test data} \label{sbsbsec:data}
    We conducted numerical experiments of HRTF interpolation to quantitatively and qualitatively evaluate the effectiveness of the proposed method.
            To generate training, validation, and test data, we used the HUTUBS dataset~\cite{Brinkmann+19a,Brinkmann+19b}. 
            It includes 
            HRIRs
            of 94 subjects, excluding duplicates. 
            The HRIRs of each subject were measured at 440 
            measurement positions
            on a sphere of radius $r=\SI{1.47}{m}$.
            %
            We used 77, 10, and 7 subjects to generate training, validation, and test data, respectively.
            %
            The HRTFs were obtained from the HRIRs as follows. First, the HRIRs were resampled at $\SI{32}{kHz}$ and the filter length was set to $256$ by zero padding. Second, FFT at $256$ points was performed, and after taking the complex conjugate, only the positive frequency bins were extracted.
            Finally, HRTFs at $128$ frequency bins $\SI{125}{Hz},\SI{250}{Hz},\ldots,\SI{16}{kHz}$ were obtained.
            
    \subsubsection{Compared methods} \label{sbsbsec:compared}
            We conducted a preliminary experiment and compared an HRTF interpolation method based on a shallow NN~\cite{Lemaire+05} with the RLR-based method.
            We found that the RLR-based method greatly outperformed the shallow-NN-based method in LSD, and we chose the RLR-based method for comparison.
            %
            The expansion coefficients were obtained up to the order of $\min\{\left\lceil \NP k_l R/2\right\rceil,N \}$ for each of the cases where 
            $B^\prime=9,16,\ldots,196$
            for the subjects in the test data.
            Here, $\NP$ is Napier's constant and $R=\SI{0.45}{m}$.
            The $B^\prime$ 
            measurement positions
            were obtained by sampling the nearest neighbor points of the points contained in the spherical $t$-design~\cite{Chen&Womersley06} with $t=\sqrt{B^\prime}-1$ from the total 440 
            measurement positions.
            The HRTFs at 440 
            target positions
            were obtained from the estimated expansion coefficients.
            As an evaluation metric, LSD in \eqref{eq:LSD} was calculated.
            For the regularization parameter $\lambda$ and the maximum truncation order $N$, the four settings shown in Table~\ref{table:lininv-setting} were used.
            \begin{table}[tb]
              \centering
              \caption{Setting of the RLR-based method~\cite{Duraiswami+04} used in the experiment. Two regularization parameters and two maximum truncation orders were used. 
              \textsf{U} and \textsf{B} in the labels indicate that the system was underdetermined and balanced, respectively.}
              \begin{tabular}{ccc}
                \hline
                Label &
                $\lambda$
                & 
                $N$
                \\ \hline
                \textsf{RLR, RP6, U} & $10^{-6}$ & $19$ \\
                \textsf{RLR, RP6, B} & $10^{-6}$ & $\sqrt{B^\prime}-1$ \\
                \textsf{RLR, RP7, U}  & $10^{-7}$ & $19$ \\
                \textsf{RLR, RP7, B}  & $10^{-7}$ & $\sqrt{B^\prime}-1$ \\
                \hline
              \end{tabular}
              \label{table:lininv-setting}
            \end{table} 
            
            %
            For the proposed method, the logarithmic magnitudes of measured HRTFs were standardized to have zero mean and a unit variance before fed into the encoder.
            The reverse operation of the above standardization was applied to the decoder outputs.
            The mean and variance were determined using the training data. 
            %
            
            The proposed model was trained with an Adam optimizer for 1000 epochs. The learning rate was set at $10^{-3}$.
            The LSD for the validation data was calculated for each epoch, and the model with the lowest validation LSD was finally selected.
            During training and validation, $B^\prime$ and $B$ were set to 440.
            For the test data, we varied $B^\prime$ from $9$ to $196$.
            The hyperparameter $\alpha$ was set to one.
            %
 
    \subsection{Results and discussion} \label{sbsec:result}
        \begin{figure}[t]
          \centering
          \includegraphics[width=0.95\hsize]{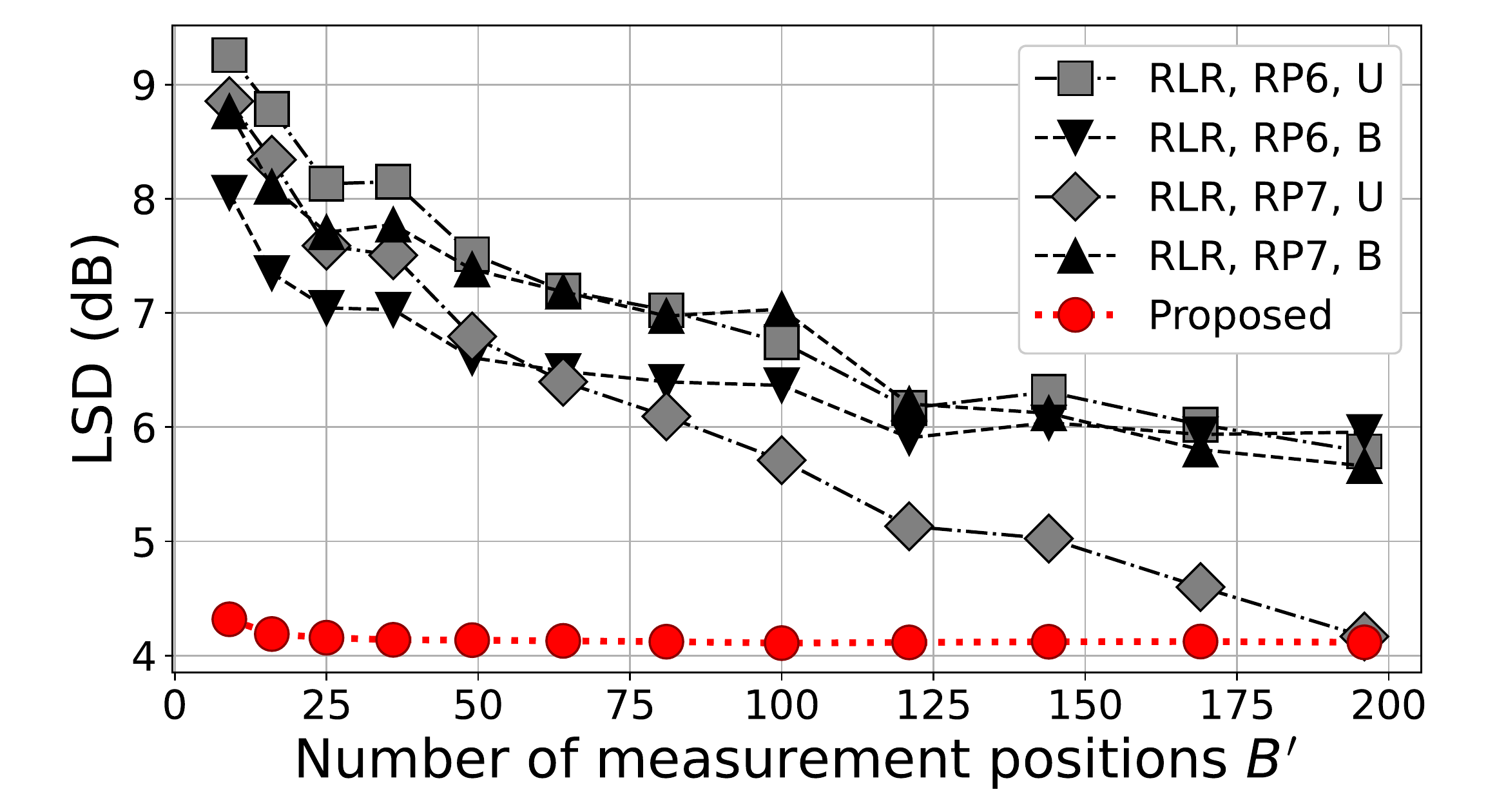} 
          \caption{
              LSDs of the proposed and RLR-based methods at various numbers of measurement positions.
              \textsf{Proposed} denotes the proposed method and the remaining labels are summarized in Table~\ref{table:lininv-setting}.
          }
          \label{fig:LSD}
        \end{figure}
        \begin{figure}[t]
          \centering
          \includegraphics[width=1.0\hsize]{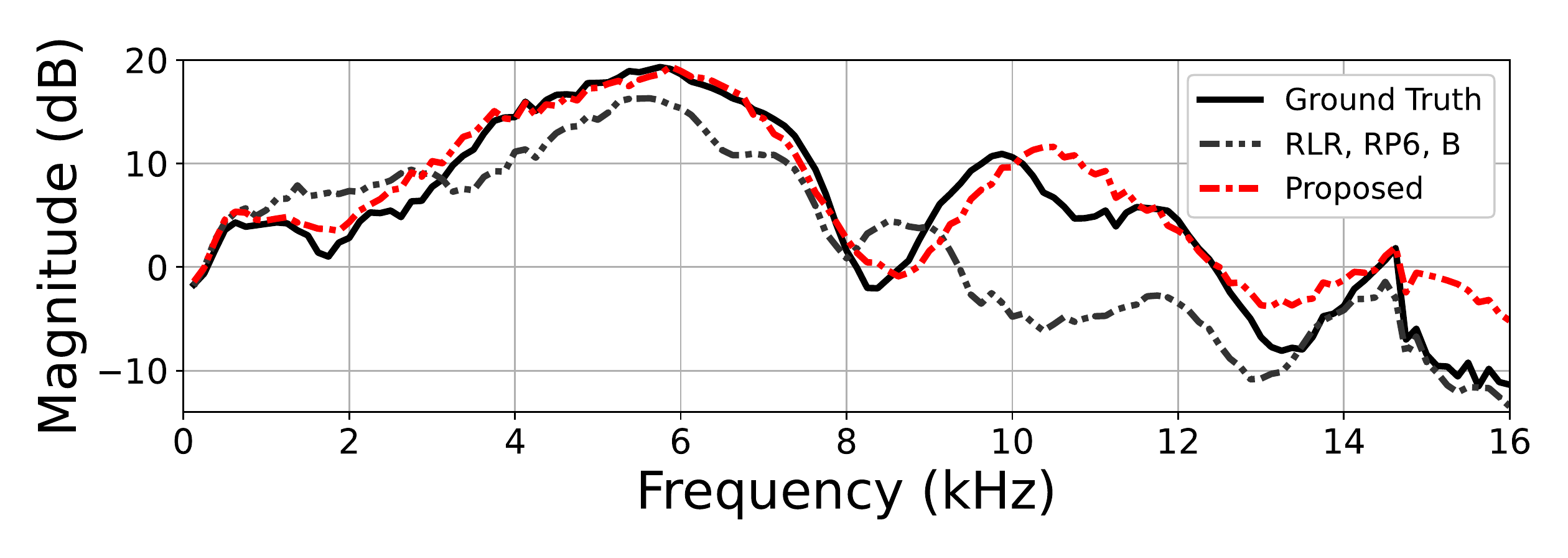} 
          \caption{
            Example of magnitude frequency responses of ground truth and estimated HRTFs with $B^\prime=9$.
            \textsf{Ground Truth} and \textsf{Proposed} denote the ground truth and the proposed method, respectively, and \textsf{RLR, RP6, B} is defined in Table~\ref{table:lininv-setting}.
          }
          \label{fig:mag-test}
        \end{figure}
        Fig.~\ref{fig:LSD} shows the LSD of HRTFs at 440 
        target positions
        obtained by 
        the proposed and RLR-based methods.
        We can see that the LSDs of the RLR-based methods increased as the number of measurement positions $B^\prime$ decreased, i.e., the accuracy decreased, regardless of whether the system was underdetermined or balanced.
        %
        On the other hand, 
        the LSDs of the proposed method were almost similar regardless of $B^\prime$. 
        Furthermore, the proposed method can interpolate HRTFs at only $B^\prime=25$ with almost the same accuracy as the RLR-based method at $B^\prime=196$.
        Note that the proposed method does not focus on the situation where HRTFs at a large number of 
        measurement positions
        are known, although it is expected that the accuracy of the RLR-based method will exceed that of the proposed method when the number of 
        measurement positions
        $B^\prime$ increases to more than $196$.
        
        Fig.~\ref{fig:mag-test} shows 
        magnitude frequency responses
        of HRTFs estimated by the RLR-based method and the proposed method for $B^\prime=9$. 
        The estimated HRTF by the RLR-based method had a greatly different spectral shape from the ground truth. 
        For example, although the ground truth had the peak in the frequency band between 10 to \SI{12}{kHz}, the estimate of the RLR-based method did not have such a peak.
        %
        In contrast, it can be confirmed that 
        the HRTF estimated by the proposed method appropriately captured the peaks and notches of the ground truth,
        even though the HRTF is from a subject not included in the training data.
        This indicates that the proposed method can interpolate HRTFs from a small number of 
        measurement positions
        more accurately than the RLR-based method.

\section{Conclusion}\label{sec:conclusion}
    We proposed a method of HRTF interpolation from sparsely measured HRTFs using an autoencoder with source position conditioning.
    The proposed architecture was designed from an analogy between the RLR-based method and an autoencoder architecture.
    Inspired by this analogy, we made the layers of the encoder and decoder dependent on source positions.
    To further reduce this dependence, we introduced a loss function based on cosine distances between the latent variables at different measurement positions.
    Numerical experiments showed that the proposed method can work well for subjects not included in the training data and 
    achieve an interpolation performance with only one-eighth measurements comparable to that of the RLR-based method.


\section{Acknowledgements}
This work was supported by JSPS KAKENHI Grant Number JP22H03608 and JST FOREST Program (Grant Number JPMJFR216M, Japan).


\end{document}